\documentclass[epj]{svjour}
\usepackage{latexsym}
\usepackage{graphics}
\def\slashchar#1{\setbox0=\hbox{$#1$}
   \dimen0=\wd0 \setbox1=\hbox{/} \dimen1=\wd1
   \ifdim\dimen0>\dimen1 \rlap{\hbox to \dimen0{\hfil/\hfil}} #1
   \else  \rlap{\hbox to \dimen1{\hfil$#1$\hfil}} / \fi}
\begin{document}
\title{S=--1 Meson-Baryon Scattering in Coupled Channel Unitarized
 Chiral Perturbation Theory \thanks{\emph{Contributed talk at QNP2002,
 9-14 June 2002, KFZ J\"ulich (Germany)}. Presented by E. Ruiz Arriola} }
\author{C. Garc{\'{\i}}a-Recio \inst{1}, J. Nieves \inst{1},
 \underline{E. Ruiz Arriola} \inst{1} \and M. Vicente Vacas \inst{2}
}
%
%
\institute{Departamento de F{\'{\i}}sica Moderna,  
Universidad de Granada. 18071-Granada (Spain) 
\and 
Departamento de F\'{\i}sica Te\'orica and IFIC, 
Centro Mixto Universidad de Valencia-CSIC, 
Ap. Correos 22085, E-46071 Valencia (Spain)}
\date{Received: date / Revised version: date}
%
\abstract{ The $s-$wave meson-baryon scattering amplitude is analyzed
for the strangeness $S=-1$ and isospin $I=0$ sector in a
Bethe-Salpeter coupled channel formalism incorporating Chiral
Symmetry. Four two-body channels have been considered: $\bar K N$,
$\pi \Sigma $, $\eta \Lambda $, $ K \Xi$. The needed two particle
irreducible matrix amplitude is taken from lowest order Chiral
Perturbation Theory in a relativistic formalism. Off-shell behaviour
is parameterized in terms of low energy constants, which outnumber
those assumed in previous works and provide a better fit to the
data. The position of the complex poles in the second Riemann sheet of
the scattering amplitude determine masses and widths of the $\Lambda
(1405)$ and $\Lambda(1670)$ resonances which compare well with
accepted numbers.
} 
%
\titlerunning{$S=-1$ Meson-Baryon Scattering ...}
\maketitle
\section{Introduction}
\label{intro}
The existence of baryon resonances is a non-perturbative feature of
intermediate energy QCD.  In addition to the standard relativistic
invariance, chiral symmetry (CS) and unitarity prove extremely
convenient tools to deal with this problem. In this energy range,
hadronic degrees of freedom seem to be the relevant ones in terms of
which the symmetries may be easyly incorporated~\cite{Pich95}.  Heavy
Baryon Chiral Perturbation Theory (HBChPT)~\cite{JM91,BK92}
incorporates CS at low energies in a systematic way, and has provided
a satisfactory description of $\pi N$ scattering in the region around
threshold~\cite{Mo98,fms98,fm00} but fails to reproduce the resonance
region. The $s-$wave meson-baryon scattering for the strangeness
$S=-1$ and isospin $I=0$ sector incorporating CS and unitarization has
been studied in previous
works~\cite{KSW95,OR98,Kr98,OM01,Ke01,ORB02,LK02}. The need for
unitarization in this reaction becomes obvious after the work of
Ref.~\cite{Ka01} where it is shown that HBChPT to one loop fails
completely in the $\bar K N $ channel already at threshold due to
nearby subthreshold $\Lambda (1405) $-resonance.

We report here on results obtained for the s-wave $S=-1$ , $I=0$ meson
baryon reaction in a Bethe-Salpeter-Equation (BSE) coupled channel
approach, extending the works of Refs.~\cite{EJ99,JE01,JE01b}.  We
also improve on a previous approach~\cite{OR98} by reparameterizing
off-shell effects as low energy constants, in the spirit of an
Effective Field Theory. We consider four coupled channels: $\pi
\Sigma$, $\bar K N$, $\eta \Lambda$ and $K \Xi$ and take into account
$SU(3)-$breaking symmetry effects but assume isospin symmetry. More
details can be found in Ref.~\cite{CJ02}.

\section{Theoretical framework} \label{sec:thf}
The coupled channel scattering amplitude for the baryon-meson process
in the isospin channel $I=0$ is given by
\begin{equation} 
T_P = \bar u_B ( P-k', s_B) t_P (k,k') u_A
 (P-k,s_A)\label{eq:deftpeque}
\end{equation} 
Here, $u_A (P-k, s_A)$ and $u_B (P-k', s_B)$ are baryon Dirac spinors
normalized as ${\bar u }u = 2M$, $P$ is the conserved total CM four
momentum, $P^2 =s$, and $t_P (k,k') $ is a matrix in the Dirac and
coupled channel spaces. Further details on normalizations and
definitions of the amplitudes can be seen in Ref.~\cite{JE01b}. To
evaluate the amplitude $t_P$ we solve the BSE
\begin{eqnarray} 
t_P (k,k') &=& v_P (k,k') \nonumber \\ + &{\rm i}& \int { d^4 q \over
(2\pi)^4 } t_P (q,k') \Delta(q) S(P-q) v_P (k,q) \label{eq:bse}
\end{eqnarray}
where $t_P( k,k')$ is the scattering amplitude defined in
Eq.~(\ref{eq:deftpeque}), $v_P(k,k')$ the two particle irreducible
Green's function (or {\it potential} ), and $ S(P-q)$ and $\Delta (q)
$ the baryon and meson exact propagators respectively. The above
equation turns out to be a matrix one, both in the coupled channel and
Dirac spaces. For any choice of the {\it potential} $v_P(k,k')$, the
resulting scattering amplitude $t_P( k,k')$ fulfills the coupled
channel unitarity condition, discussed in Eq.~(21) of
Ref.~\cite{JE01b}.  The BSE requires some input potential and baryon
and meson propagators to be solved. At lowest order of the BSE-based
chiral expansion~\cite{EJ99}, we approximate the iterated {\it
potential} by the chiral expansion lowest order meson-baryon
amplitudes in the desired strangeness and isospin channels, and the
intermediate particle propagators by the free ones (which are diagonal
in the coupled channel space). From the meson-baryon chiral
Lagrangian~\cite{Pich95} (see Sect. IIA of Ref.~\cite{JE01b}), one
gets at lowest order for the {\it potential}:
\begin{equation}
v_P (k,k') = t_P^{(1)} (k,k') = {D \over f^2} ( \slashchar{k}+\slashchar{k}' ) 
\end{equation}
with $D$ the coupled-channel matrix,
\begin{eqnarray} 
\nonumber \\ 
\phantom{D = \frac14} && \matrix{    \bar K N     \, \, & \quad \pi \Sigma \, \, \,& \quad  \eta
\Lambda  & \quad K \Xi  & \quad \qquad \qquad  } \nonumber  
\\ 
D = \frac14 && \left( \matrix{
 -3  &   \sqrt{\frac32}  &   -\frac3{\sqrt{2}} & 0 \cr 
 \sqrt{\frac32}  &    -4   & 0  &   -\sqrt{\frac32} \cr
 -\frac3{\sqrt{2}}  &   0  &  0  &   \frac3{\sqrt{2}} \cr
 0  &   -\sqrt{\frac32}  &  +\frac3{\sqrt{2}}  &  -3  
} \right)            
\qquad  \matrix{   \bar K N \cr \cr  \pi \Sigma \cr  \cr  \eta \Lambda
\cr \cr  K \Xi } 
\label{eq:d-matrix} 
\end{eqnarray} 
The $s-$wave BSE can be solved and renormalized up to a numerical
matrix inversion in the coupled channel space~\cite{JE01}.

\section{Numerical results} \label{sec:nr}

We use the following numerical values for masses and weak decay
constants of the pseudoscalar mesons (all in MeV), $ m_K = m_{\bar
K}=493.68 $, $ m_\pi = 139.57 $, $ m_\eta = 547.3 $, $ M_p = 938.27 $,
$ M_\Sigma = 1189.37 $, $ M_\Lambda = 1115.68 $, $ M_\Xi = 1318.0 $
and $ f_\pi = f_\eta = f_K = 1.15\times 93.0 $ where for the weak
meson decay constants we take for all channels an averaged value.

\begin{figure}
\resizebox{0.5\textwidth}{!}{%
  \includegraphics{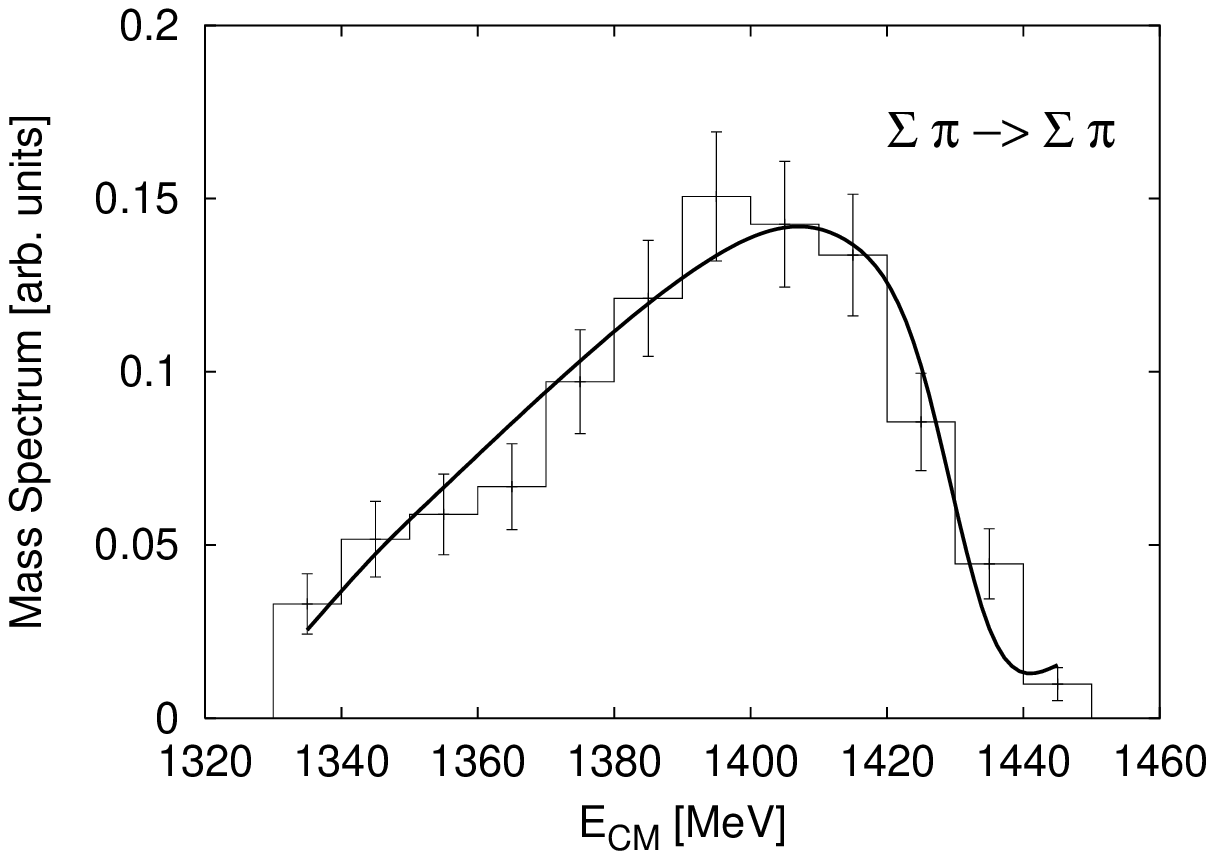}\includegraphics{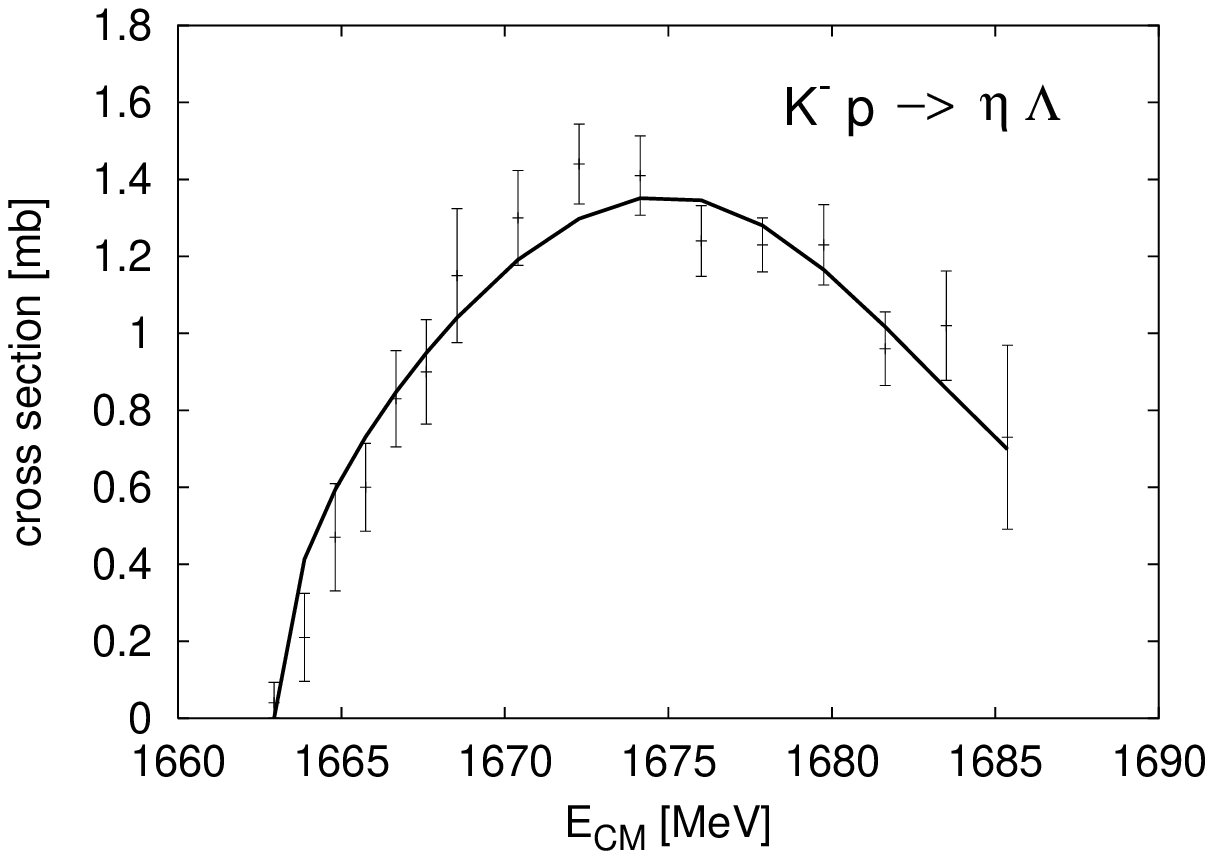}
}
\resizebox{0.5\textwidth}{!}{%
  \includegraphics{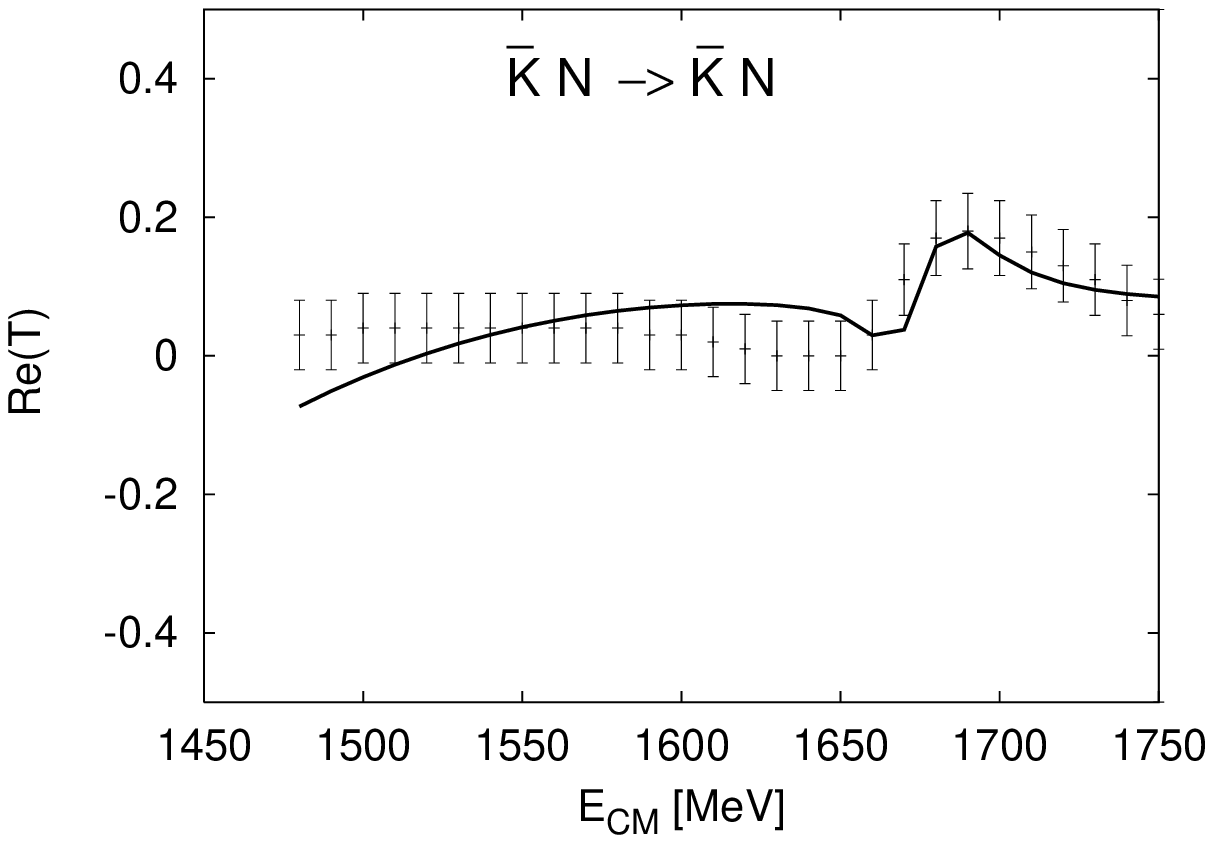}\includegraphics{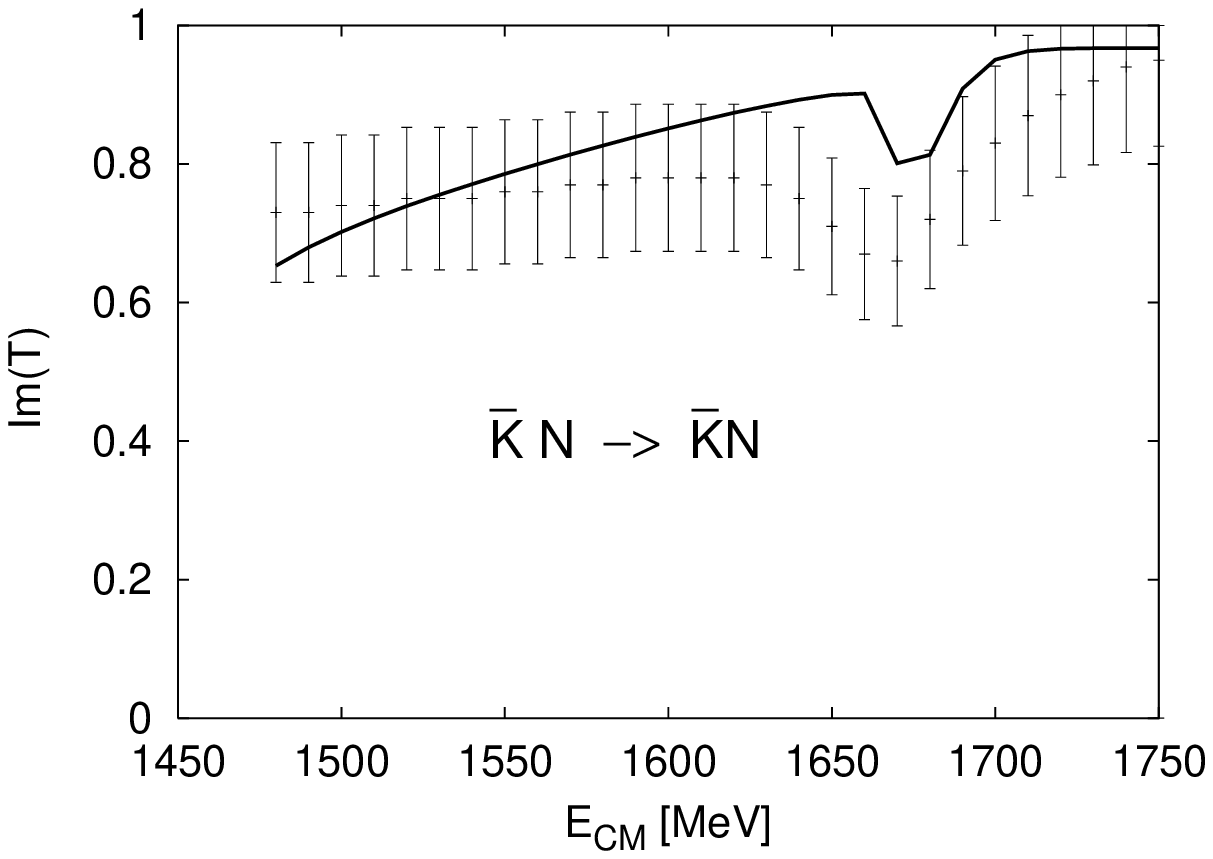}
}
\resizebox{0.5\textwidth}{!}{%
  \includegraphics{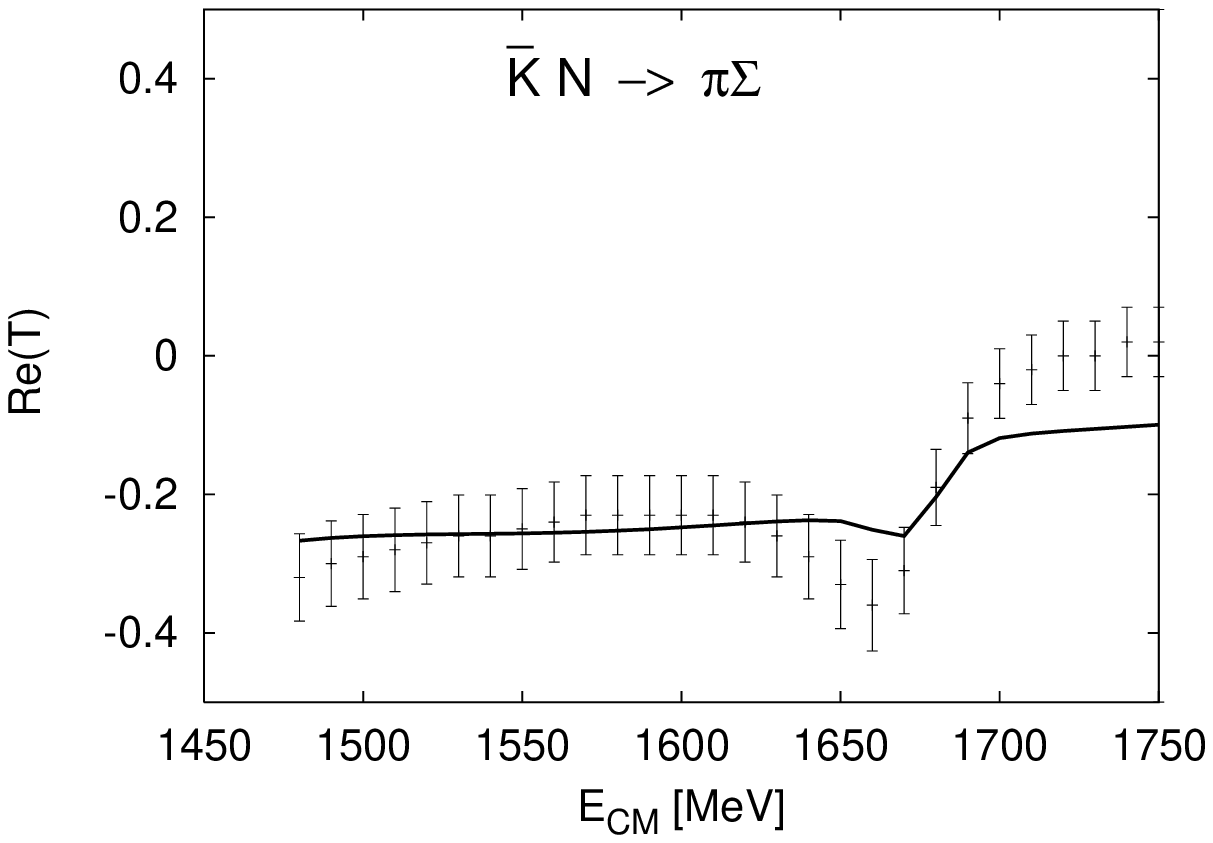}\includegraphics{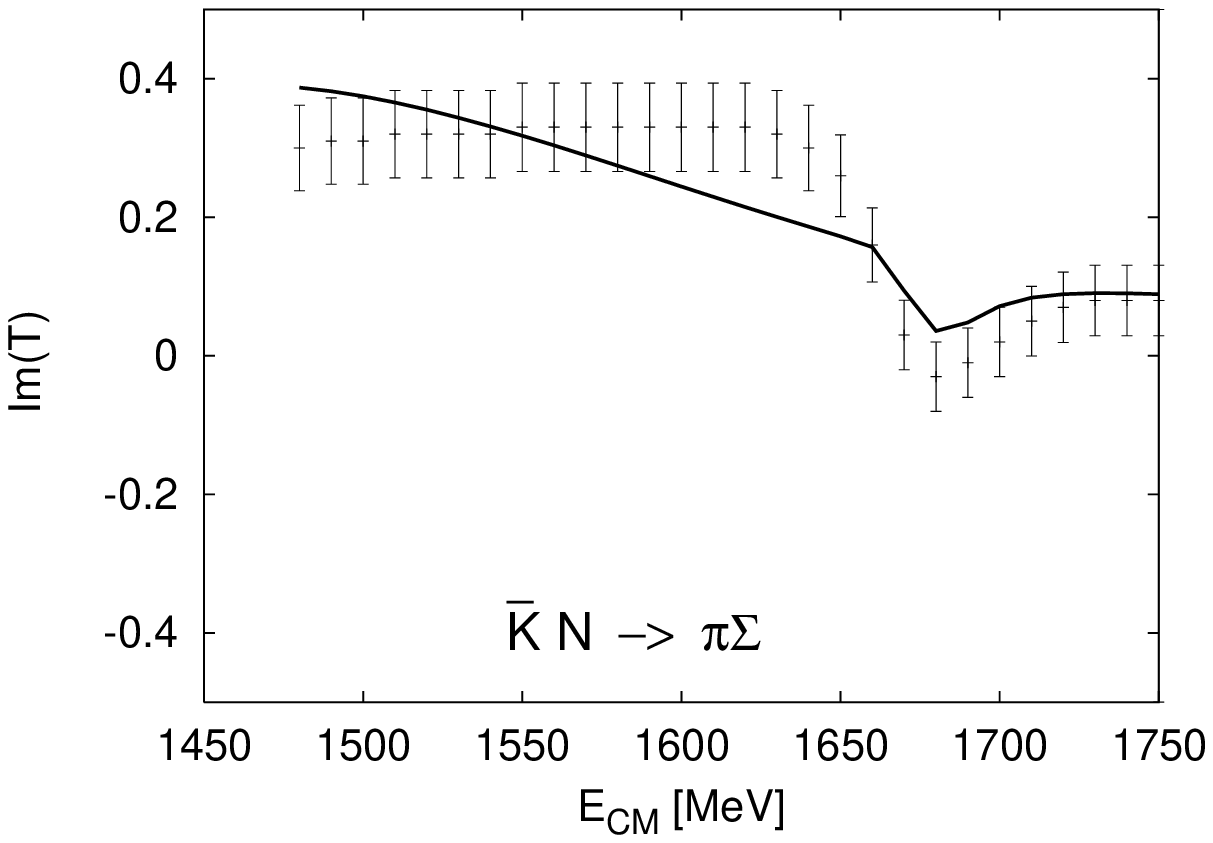}
}
\caption{Best fit results for the Bethe-Salpeter equation in the
$S=-1$, $I=0$ channel (solid lines). {\bf Upper pannel:} Experimental
data for $\pi\Sigma\to\pi\Sigma$ and $K^- p\to \eta\Lambda$ are from
Refs.~\protect\cite{He84}~and~\protect\cite{St01}, respectively. {\bf
Middle pannel:} The real (left panel) and imaginary (right panel)
parts of the $s-$wave $T-$matrix, with normalization specified in the
main text, for elastic $\bar K N \to \bar K N$ process in the $I=0$
isospin channel as functions of the CM energy.  Experimental data are
taken from the analysis of Ref.~\protect\cite{Go77} with the errors
stated in the main text. {\bf Lower Pannel:} Same as middle pannel for
the inelastic channel $\bar K N \to \pi \Sigma$.}
\label{fig:1}       
\end{figure}

\subsection{ Fitting procedure} 
\label{sec:data}
We perform a $\chi^2-$fit, with 12 free parameters,  to the following
set of experimental data and conditions:
\begin{enumerate}
\item $S_{01}(L_{2T2J})$ $\bar K N \to \bar K N$ and $\bar K N \to \pi
\Sigma$ scattering amplitudes (real and imaginary parts)~\cite{Go77}
in the CM energy range of $1480 \, {\rm MeV} \le \sqrt{s} \le 1750 \,
{\rm MeV}$. In this CM energy region, there are a total number of 56
data points (28 real and 28 imaginary parts) for each channel. The
normalization used in Ref.~\cite{Go77} is different of that used here
and their amplitudes, $T_{ij}^{{\rm Go77}}$, are related to ours by:
$
T_{ij}^{{\rm Go77}}={\rm sig}(i,j)|\vec{k}_i|~\left[ f_0^\frac12 (s)
\right]_{j \leftarrow i} ,
$
where sig$(i,j)$ is $+1$ for the elastic channel and $-1$ for the
$\bar K N \to \pi \Sigma$ one.  On the other hand, and because in
Ref.~\cite{Go77} errors are not provided, we have taken for those
amplitudes errors given by
%
$ \delta T_{ij}^{{\rm Go77}}=\sqrt{(0.12~T_{ij}^{{\rm Go77}})^2+0.05^2}
$
%
in the spirit of those used in Ref.~\cite{Ke01}.
\item $S_{01}-\pi \Sigma$ mass spectrum ~\cite{He84}, $1330 \, {\rm
MeV} \le \sqrt{s} \le 1440 \, {\rm MeV}$. In this CM energy region,
there are a total of thirteen 10~MeV bins and the experimental data
are given in arbitrary units. To compare with data, taking into
account the experimental acceptance of 10~MeV, we compute:
\begin{eqnarray}
\Delta\sigma / \Delta [M_{\pi\Sigma}({\rm i})] = C \int_{M^-}^{M^+}
\left|\left [f^\frac{1}{2}_0(s=x^2)\right]_{2\leftarrow 2}\right|^2
\nonumber \\  \times |\vec{k}_2(s=x^2)|~x^2~dx
\end{eqnarray}
where $C$ is an arbitrary global normalization factor\footnote{We fix
it by setting the area of our theoretical spectrum,
$\sum_i\frac{\Delta\sigma}{\Delta[ M_{\pi\Sigma}({\rm i})]}$, to the
total number of experimental counts $\sum_i N_i$.}, $ M^{\pm}
=M_{\pi\Sigma}({\rm i}) \pm 5{\rm MeV} $ and $i$ denotes the bin with
central CM energy $M_{\pi\Sigma}(i)$. Hence, there are only 12
independent data points. Finally, we take the error of the number of
counts, $N_i$, of the bin $i$ to be $1.61\sqrt{N_i}$ as in
Ref.~\cite{Da91}.

\item The $K^- p \to \eta \Lambda$ total cross section of
Ref.~\cite{St01}, \newline $1662 {\rm MeV} \le \sqrt{s} \le 1684 {\rm MeV}$. We
use the Crystal Ball Collaboration precise new total cross-section
measurements (a total of 17 data points compiled in Table I of
Ref.~\cite{St01}) for the near-threshold reaction $K^-p \to
\eta\Lambda$, which is dominated by the $\Lambda(1670)$ resonance.  We
assume, as in Ref.~\cite{St01}, that the $p-$ and higher wave
contributions do not contribute to the total cross-section.
\end{enumerate} 

Finally, we define the $\chi^2$, which is minimized, as 
\begin{equation}
\chi^2/N_{\rm tot} = \frac{1}{N} \sum_{\alpha=1}^{N}
\frac{1}{n_\alpha}\sum_{j=1}^{n_\alpha}
\left( \frac {x_j^{(\alpha )\,{\it th}}-x_j^{(\alpha
  )}}{\sigma_j^{(\alpha)}}\right)^2\,  ,
\end{equation}
where $N=4$ stands for the four sets of data used and discussed
above. Though we have considered four coupled channels, three-body
channels, for instance the $\pi\pi\Sigma$ one, are not explicitly
considered, as it has been also assumed previously in
Refs.~\cite{ORB02} and~\cite{St01}.
\begin{table}
\caption{Resonance Masses and Widths (in MeV)}
\label{tab:1}       
\begin{tabular}{clll}
\hline\noalign{\smallskip} \phantom{pepepepe}& first & second & third
\phantom{pepe} \\
\noalign{\smallskip}\hline\noalign{\smallskip} $M_R$ & $1368 \pm 12$ &
$1443 \pm 3$ & $ 1677.5 \pm 0.8$ \\ $\Gamma_R $ & $250 \pm 20$ & $50
\pm 7$ & $
29.2 \pm 1.4 $\\
\noalign{\smallskip}\hline
\end{tabular}
\end{table}
\subsection{Poles and Couplings} 
Resonances are defined as poles in the second Riemann sheet of the
$s-$complex plane. Around them, the scattering matrix behaves as
\begin{eqnarray}
\left[t(s)\right]_{ij} & \to & \frac{ 2 M_R  g_i g_j }{s-M_R^2 + {\rm i}
M_R  \Gamma_R },
\label{eq:tpole}
\end{eqnarray}
We find three poles in the Second Riemann Sheet which positions are
given in table~\ref{tab:1}.  Errors have been transported from those
in the best fit parameters~\cite{CJ02}, taking into account the
existing statistical correlations through a Monte--Carlo simulation.
As can be seen from Fig.~\ref{fig:2} besides the three poles appearing
in the Second Riemann Sheet, unphysical poles show up in the physical
sheet out of the real axis, but they do not influence the
scattering. Our resonances are not of Breit-Wigner form. For the
$\Lambda(1670)$ resonance, branching ratios, as defined in
Ref.~\cite{CJ02}, are
\begin{equation}
B_{\bar K N} = 0.24, \qquad B_{ \pi \Sigma} = 0.08, \qquad B_{ \eta \Lambda} = 0.68   \qquad 
\end{equation}
These values reasonably agree to the values 
quoted in the PDG  
(
$B_{\bar K N}= 0.20\pm 0.05, ~ B_{ \pi \Sigma}= 0.40\pm 0.20,~ B_{
 \eta \Lambda}= 0.25\pm 0.10$
)
and in Ref.~\cite{Ma02}
(
$B_{\bar K N}= 0.37\pm 0.07, ~ B_{ \pi \Sigma}= 0.16\pm 0.06,~ B_{
 \eta \Lambda}= 0.39\pm 0.08,~B_{\pi\Sigma(1385)}=0.08\pm 0.06$
). 

\vskip.3cm {\it This work has been supported by the DGES grants
no. BFM200-1326 and PB98-1367 and the Junta de Andaluc{\'\i}a.}
%
%
%
%
%

\end{document}